\documentclass[aps,prd,english,superscriptaddress, longbibliography,11pt,notitlepage, nofootinbib]{revtex4}

	\usepackage[colorlinks=true, a4paper=true, pdfstartview=FitV,
linkcolor=blue, citecolor=blue, urlcolor=blue]{hyperref}

\usepackage{amsmath}
\usepackage{amssymb}
\usepackage{graphicx}
\usepackage{hhline}
\usepackage{mwe}
\usepackage{amsbsy}
\usepackage{textcomp}
\usepackage{commath}
\usepackage{subfig}
\usepackage{slashed}
\usepackage{natbib}
\usepackage{xcolor}

\usepackage{stackengine}
\stackMath

\makeatletter
\usepackage{babel}
\numberwithin{equation}{section}
\begin{document}
\immediate\write16{<<WARNING: LINEDRAW macros work with emTeX-dvivers
                    and other drivers supporting emTeX \special's
                    (dviscr, dvihplj, dvidot, dvips, dviwin, etc.) >>}

\title{Resonance with quasinormal modes in long-range kinks' collisions}
\author{J. G. F. Campos}
\email{joao.gfcampos@upe.br}
\affiliation{Física de Materiais, Universidade de Pernambuco, Rua Benfica, 455, Recife - PE - 50720-001, Brazil}
\author{A. Mohammadi}
\email{azadeh.mohammadi@ufpe.br}
\affiliation{Departamento de Física, Universidade Federal da Pernambuco, Av. Prof. Moraes Rego, 1235, Recife - PE - 50670-901, Brazil}
\affiliation{Department of Applied Mathematics and Theoretical Physics, University of Cambridge,\\
Wilberforce Road, Cambridge CB3 0WA, United Kingdom}
\author{T. Romanczukiewicz}
\email{tomasz.romanczukiewicz@uj.edu.pl}
\affiliation{Jagiellonian University, Krakow, Poland}

\begin{abstract}

We consider a rational scalar field model in (1+1)-dimensions where the long-range character of the kinks is controllable. We show via numerical simulations that kinks with long-range tails on both sides can exhibit resonance windows. The resonant energy exchange mechanism occurs via the excitation of quasinormal modes, which we obtain via a spectral analysis. Additionally, we locate a resonance window in a family of $\phi^{10}$ models with long-range tails on both sides. Moreover, we propose a new algorithm for initializing long-range kink collisions, based on convection–diffusion dynamics.

\end{abstract}

\maketitle

\section{Introduction}
\label{intro}

Solitons are localized, stable solutions to nonlinear field equations with broad applications across physics, including condensed matter, cosmology, and field theory \cite{rajaraman1982solitons,vilenkin1994cosmic,manton2004topological}. Kinks, which are topological solitons in $1+1$ dimensions, can have rich mathematical structures, especially when integrable. However, non-integrable ones display far richer dynamical behaviors and possible outcomes when interacting with other kinks. 
In particular, in kink-antikink interactions, possible outcomes depend on the initial velocity. At low velocities, kink and antikink pairs may annihilate, with or without forming a long-lived oscillatory bound state known as a bion \cite{campbell1983resonance,campos2021interaction}. Above a critical velocity, they have an inelastic scattering, producing some radiation. Remarkably, in an intermediate regime, the outcome can alternate chaotically between annihilation and escape, giving rise to so-called resonance windows. The latter happens when there is a possibility for the kink-antikink to store the initial kinetic energy in vibrating internal modes during the collision and later convert back into kinetic energy, allowing the kinks to separate.
There are several known possible energy exchange mechanisms in the literature responsible for the appearance of resonance windows. Among them are the lowest energy vibrational bound mode (BM) of individual kinks \cite{campbell1983resonance,campbell1986kink}, bound vibrational mode of the kink-antikink pair \cite{dorey2011kink},  a sphaleron \cite{adam2021sphalerons}, or even a fermion field \cite{bazeia2022resonance}.

Besides the possibilities mentioned above, it has been shown that quasinormal modes (QNMs) can mediate the energy exchange responsible for resonance windows, despite their intrinsic energy leakage \cite{dorey2018resonant,campos2020quasinormal}. In this work, we investigate the same phenomenon in a more extreme setting for kinks with long-range tails from both sides, dubbed as double long-range kinks. In this case, vibrational bound modes are absent, and QNMs are the only possibility to store energy and present resonance windows.

Initiating collisions involving long-range kinks presents significant challenges. A simple additive ansatz fails due to the kinks' highly interactive character.  In the last few years, some more elaborate approaches have appeared to tackle this problem. In particular, in \cite{christov2019long}, the authors employed the split-domain (SD) ansatz in kink-antikink collisions. This method performs reasonably well for kinks that are softly long-range. More accurate methods were suggested in \cite{christov2019long,christov2021kink,campos2021interaction} for kinks initially at rest and moving with a nonzero velocity. The  general method entails two layers of minimization, for the kinks' initial configuration and their velocity fields. It is proven to be very precise, however, computationally costly. 
In \cite{campos2024collisions}, the authors proposed a much more efficient method, using a kink on an impurity, with a precision somewhere in between SD and two layers of minimization. In this work, we introduce an alternative efficient method for initiating long-range kink collisions. Although it is slightly less precise than the two-layer minimization scheme, it is physically intuitive, as it is based on a convection-diffusion equation. The underlying idea can be naturally interpreted in terms of gradient flow dynamics.

This paper is organized as follows. In Sec.\ref{sec:mod}, we start with a toy model with the potential as a rational function, with two adjustable parameters. We identify the regimes in which the kink solutions transition from short-range to long-range behavior. Although explicit kink solutions are not available analytically, we estimate the asymptotic behavior of their tails in different parameter regimes. We also examine the spectral structure obtained from the linear stability analysis, which serves as a guide for interpreting the collision results.
In Sec. \ref{polphi10}, we conduct a similar analysis for a more realistic model with $\phi^{10}$ potential. The kinks in this model are more long-range compared to those in the rational model.
Section \ref{sec:res} is devoted to the kink-antikink collisions and the analysis of the energy transfer mechanism via QNMs for both rational and $\phi^{10}$ models. In Sec. \ref{sec:conv-diff}, we introduce the convection and diffusion algorithm and compare the accuracy with the other known methods in the literature. Finally, Sec.~\ref{sec:res} summarizes the main findings and offers concluding remarks.

\section{Rational Model}
\label{sec:mod}

We study the following scalar field theory in $(1+1)$-dimensions, defined by the Lagrangian density  
\begin{equation}
    \mathcal{L} = \frac{1}{2} \partial_\mu \phi \partial^\mu \phi - U(\phi),
\end{equation}
where the potential is a rational function given by \cite{dorey2018resonant, dorey2024oscillons}
\begin{equation}\label{rat_potential}
    U(\phi; m, \varepsilon) = V + \frac{m^2 - 4}{4} \frac{\varepsilon V}{V + \varepsilon} \, ,
\end{equation}
with \( V = \frac{1}{2}(\phi^2 - 1)^2 \) being the usual \(\phi^4\) potential.

\begin{figure}
    \centering
    \includegraphics[width=1\linewidth]{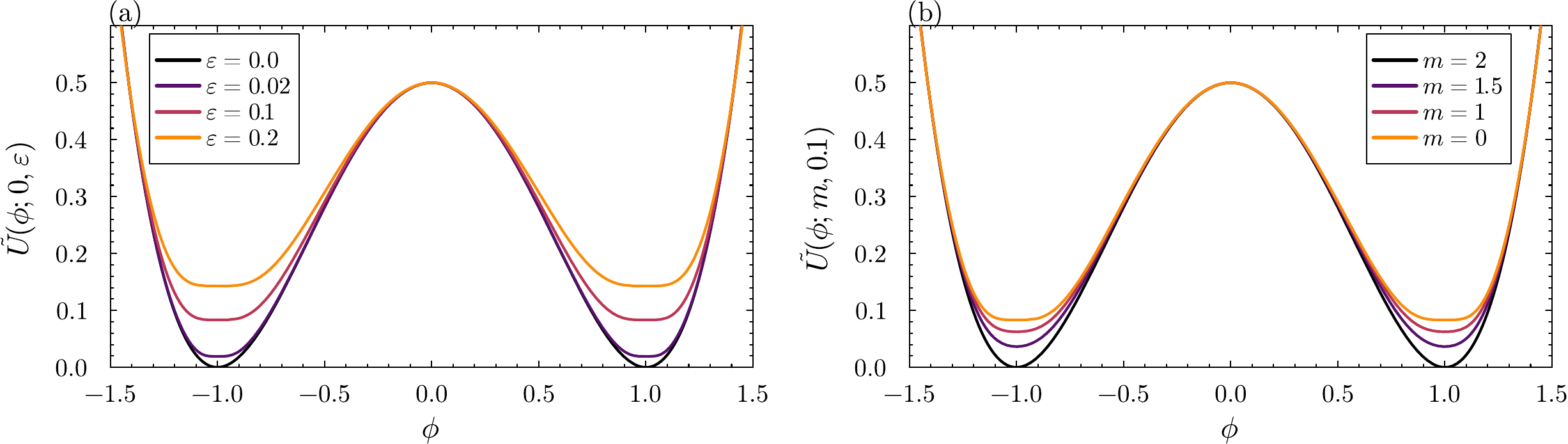}
    \caption{Potential as a function of $\phi$ for several values of $m$ and $\varepsilon$. }
    \label{fig:rat_pot}
\end{figure}

To help visualization, in Figs.~\ref{fig:rat_pot} (a) and (b), we plot the potential for several values of $m$ and $\varepsilon$, applying the following shift
\begin{equation}
    \tilde{U}(\phi; m, \varepsilon) = U(\phi; m, \varepsilon) + \frac{1}{2} - U(0; m, \varepsilon).
\end{equation}
From the curvature of the potential near its minima, one anticipates that the kink becomes more long-range as \( m \) decreases and \( \varepsilon \) increases.
In particular, the kink solutions are short-range for $V\gg \varepsilon$, regardless of the value of $m$. In this case, the potential $U(\phi;m,\varepsilon) \to V$. The transition from short-range to long-range behavior happens around $V(\Phi_K)\approx\varepsilon$ when $m=0$. More specifically, for $m=0$, 
\begin{equation}
U(\phi;0,\varepsilon)\approx \begin{cases}
V-\varepsilon+\mathcal{O}(\varepsilon^2)& \textrm{for  } V>\varepsilon \, \,  \left(\phi^2<1-\sqrt{2\varepsilon}\right)\\
\frac{V^2}{\varepsilon}+\mathcal{O}(V^3)& \textrm{for  } V<\varepsilon \, \, \left(\phi^2>1-\sqrt{2\varepsilon}\right)\end{cases}    
\end{equation}

\begin{figure}
    \centering
    \includegraphics[width=1\linewidth]{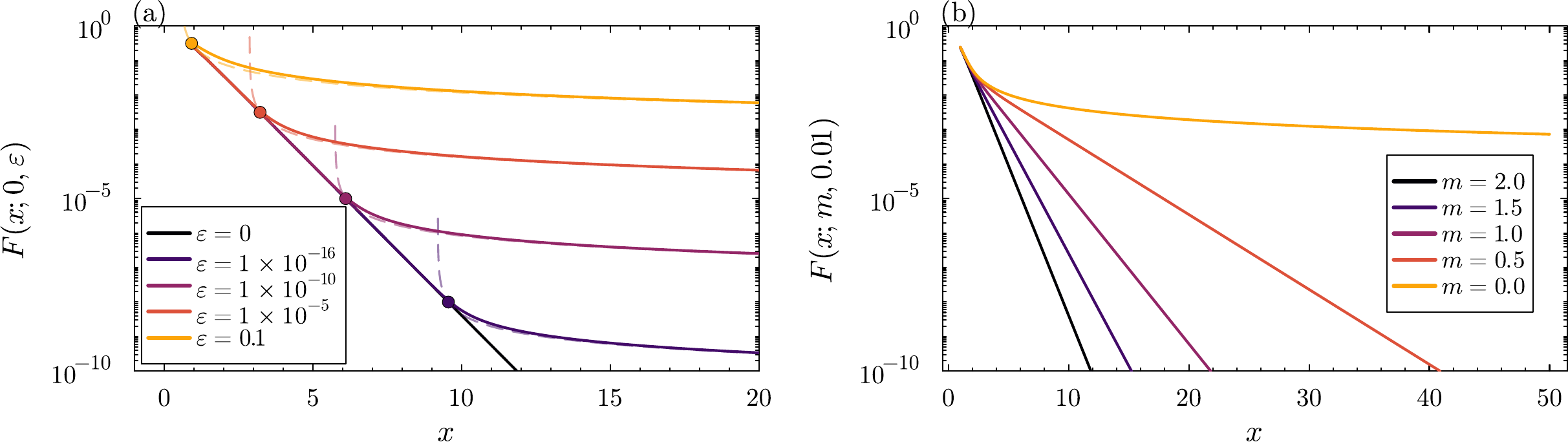}
    \caption{Kink tail on a logarithmic scale as a function of $x$ for several values of $m$ and $\varepsilon$. }
    \label{fig:rat_kinks}
\end{figure}

In the first regime, $V>\varepsilon$, in the zeroth order of $\varepsilon$ the potential is equal to the $\phi^4$ potential. In this case, the BPS equation gives the standard $\phi^4$ kink $\phi_K(x)=\tanh(x-x_0)$, where we can set the integration constant $x_0 = 0$ without loss of generality. The validity of the $\phi^4$ approximation is restricted to the region $|x| < L_s$, with
\begin{equation}
    L_s=\textrm{atanh}\left(\sqrt{1-\sqrt{2\varepsilon}}\right), 
\end{equation}
which becomes a better approximation for $\varepsilon \ll 1$. In this case, $L_s$ can be approximated by $ L_s\approx-\frac{1}{4}\ln(\varepsilon/8)$.
This feature is clearly visible in the numerically computed solutions shown in Fig. \ref{fig:rat_kinks} on a logarithmic scale, where the kink profile is defined as $\phi_K(x)=1-F(x;m,\epsilon)$. Initially, $F(x;m,\epsilon)$ follows a nearly straight line until it reaches the point $x=L_s$ when it diverges from exponential decay. 

In the second regime, $V<\varepsilon$, close to the vacuum $\phi=1$, the BPS equation simplifies to
\begin{equation}
    \phi'\approx \sqrt{\frac{2}{\epsilon}}V \left(1-V/(2\epsilon)\right)\approx \kappa F^2(1-F),
\end{equation}
with $\kappa=\sqrt{\frac{8}{\varepsilon}}$.
Integrating this equation results in a power-law tail in the form
\begin{equation}
    \phi_K(x)\approx1-\frac{1}{\kappa(x-A)}-\frac{\ln \kappa (x-A)}{\kappa^2(x-A)^2}
\end{equation}
where $A$ is an integration constant, yet to be determined. The power-law decay reflects the long-range nature of the kink in this regime. 
To determine $A$, we suppose $\varepsilon \ll 1$ in order to keep only up to $1/(x-A)$ term in the above expression. 
We assume that the exponential and power-law regimes match at the point where $V(\phi) = \varepsilon$. This condition yields
\begin{equation}
    A\approx L_s-\frac{1}{2},
\end{equation}
which matches exactly with the small circles marked in Fig.~\ref{fig:rat_kinks}(a).

\subsection{Spectral Analysis}
\label{sec:specrational}

We adopt the standard linear stability analysis to analyze the behavior of small-amplitude perturbations around the kink solution. We consider fluctuations of the form $\phi(x,t)=\phi_K(x)+\xi(x,t)$, where $\phi_K(x)$ is the static kink solution and $\xi(x,t)$ is a small perturbation. Expanding the equation of motion to linear order in $\xi$, we obtain a Schrödinger-like equation
\begin{equation}
    \xi_{tt}-\xi_{xx}+V_{lin}(x;m,\varepsilon)\xi=0,\qquad V_{lin}(x;m,\varepsilon) = \left.\frac{\partial^2 U}{\partial\phi^2}\right|_{\phi=\phi_K}.
\end{equation}
Isolating Fourier modes in the form $\xi(x,t)=e^{i\omega t}\eta(x)$, the equation reduces to an eigenvalue problem for $\eta(x)$, where $\omega$ denotes the mode frequency.

For $\varepsilon = 0$, the model reduces to the well-known $\phi^4$ theory. In this case, the linearized stability potential  takes the form of a solvable Pöschl-Teller potential
\begin{equation}
 V_{lin}(x;m,0) =   4-\frac{6}{\cosh^2x}.
\end{equation}
This potential supports two bound states: a zero mode associated with translational invariance $\omega=0$ and a discrete bound mode $\omega=\sqrt{3}$ representing the oscillating shape mode of the $\phi^4$ kink. Additionally, a resonance mode exists at the continuum threshold $\omega=2$, which is even under parity and non-normalizable. The spectrum contains no further resonances or antibound modes (aBMs). By antibound mode, we refer to a solution of the linearized equation with a real frequency and exponentially growing tails at spatial infinity, opposite to the localized bound states. Notably, the continuum scattering modes are completely reflectionless, a distinctive feature of the $\phi^4$ kink.

\begin{figure}
    \centering
    \includegraphics[width=1\linewidth]{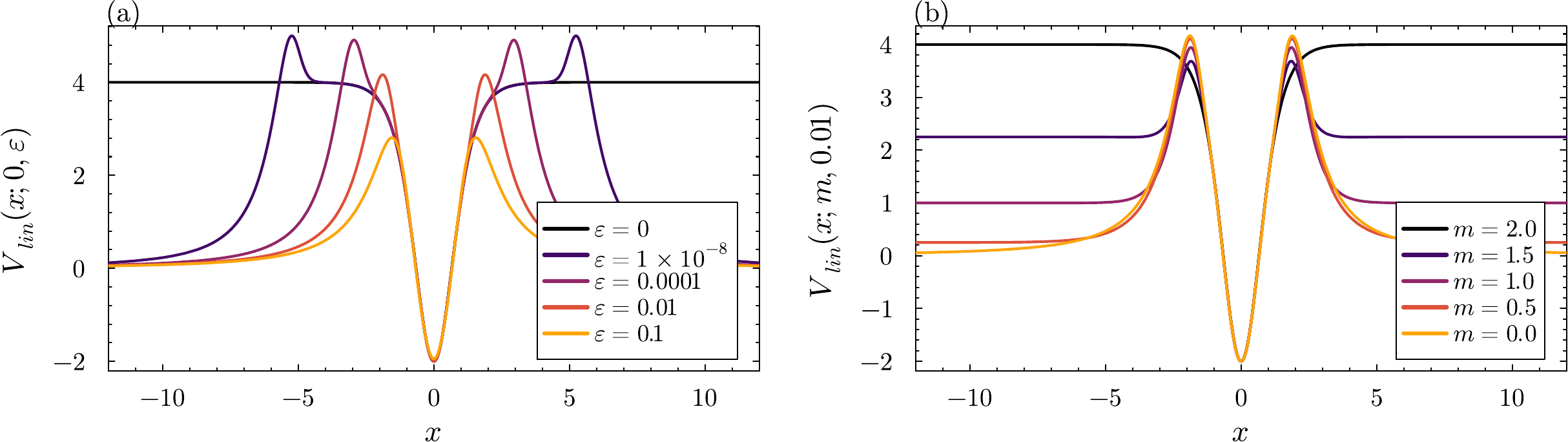}
    \caption{Stability potential (a) for fixed $m=0$ and several values of $\varepsilon$ and (b) for fixed $\varepsilon=0.01$ and several values of $m$.}
    \label{fig:rat_eff_pot}
\end{figure}

Figure \ref{fig:rat_eff_pot} shows the stability potential $ V_{lin}(x;m,\varepsilon)$ as a function of $x$ for fixed $m=0$ and several values of $\varepsilon$ and also fixed $\varepsilon=0.01$ and several values of $m$. As one can see, for $\varepsilon>0$, the potential takes a volcano shape. For the case $m=0$, it goes to zero from both sides,  reflecting the double long-range character of the kink solutions. As $\varepsilon$ grows from zero, the barriers of the potential become narrower, creating a greater leak of energy. In this case, there is no possible normalizable bound normal mode. However, there can be non-normalizable ones, like antibound, quasinormal, and threshold modes besides the continuum ones. Among these, only quasinormal modes may contribute to the energy transfer leading to resonance windows, despite the leak \cite{dorey2018resonant}. The question here is whether this can be the case for the extreme scenarios of double long-range kinks with no other possibility for the energy exchange and how long-range they can be for this to happen.

\begin{figure}
    \centering
    \includegraphics[width=1\linewidth]{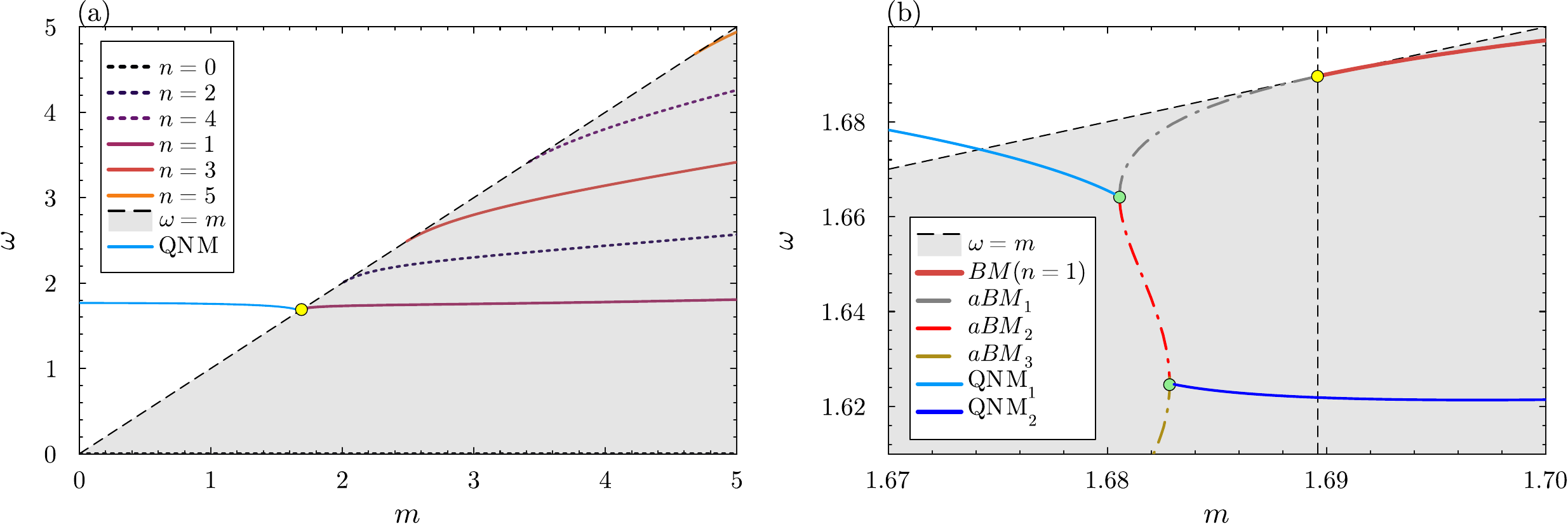}
    \caption{(a) Spectral structure (BMs) for  $\varepsilon=0.01$ along with the real part of the frequency of the lowest odd QNM. (b) The lowest odd mode, antibound modes and lowest QNMs around the threshold crossing.}
    \label{fig:rational_spectralBM}
\end{figure}

Now, let us explore the spectral structure in this model in detail. The spectral structure for fixed $\varepsilon=0.01$ and varying $m$ is shown in Figure~\ref{fig:rational_spectralBM}. As one can see, for $m>2$,there appears to be more than one bound normal mode. Notably, one bound mode emerges precisely at $m=2$, originating from the resonance at the threshold of the Pöschl-Teller potential. As $m$ decreases, the frequency of the only oscillation mode decreases gradually. At the critical value $m=m_w \approx 1.6896$
, it reaches the threshold and becomes a threshold resonance. The mode transitions into an antibound state below this point $m<m_w$. At $m=1.6806$, this antibound mode merges with another antibound mode, and for $m<1.6806$, they continue as a QNM with a complex frequency.

Quasinormal modes satisfy purely outgoing boundary conditions, which break time-reversal symmetry and the hermiticity of the linearized operator. As a result, complex eigenfrequencies are allowed. For each QNM, the complex conjugate frequency also satisfies the equations, corresponding to a purely incoming wave solution. While such incoming solutions are rarely physically relevant, their existence explains how two antibound modes can combine into a single QNM, which formally coexists with its complex-conjugate partner.

It is worth mentioning that another such bifurcation involving a QNM occurs at $m=1.6828$, where a QNM transitions into two antibound modes. One of them merges with the antibound mode originating from the vibrational mode at $m=1.6806$. This intricate structure is illustrated in detail in Fig.~\ref{fig:rational_spectralBM}(b).

\begin{figure}
    \centering
    \includegraphics[width=1\linewidth]{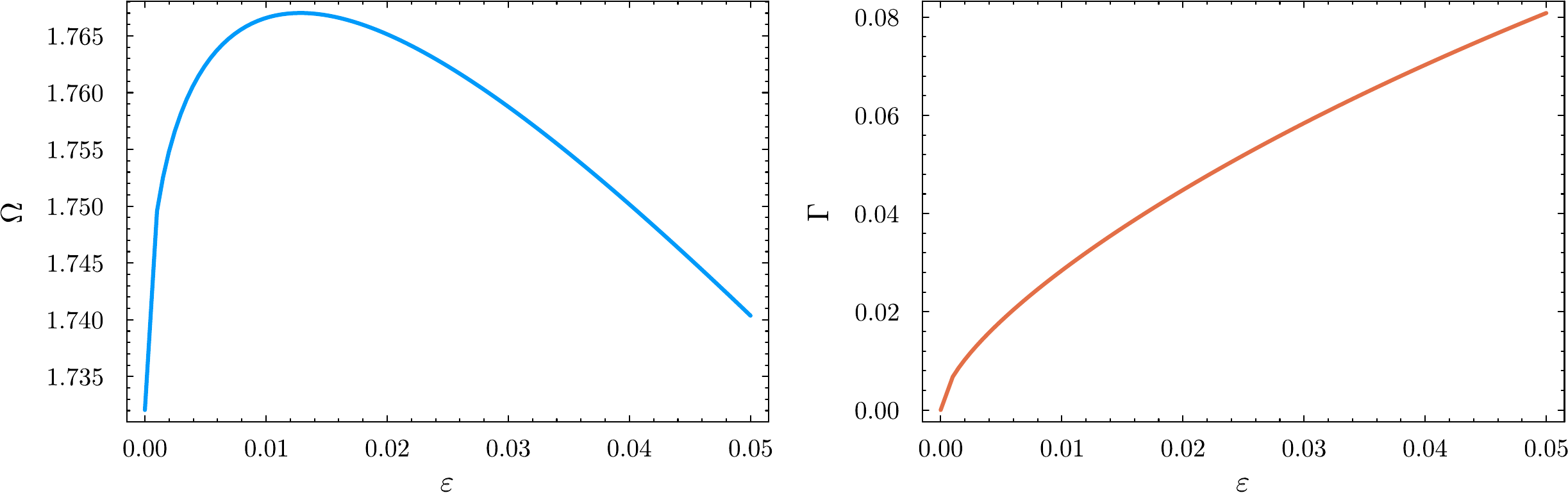}
    \caption{Spectral structure of the lowest QNM with the frequency defined as $\omega=\Omega+i\Gamma$  for $m=0$ in the rational model.}
    \label{fig:rational_spectral}
\end{figure}

\begin{figure}
    \centering
    \includegraphics[width=1\linewidth]{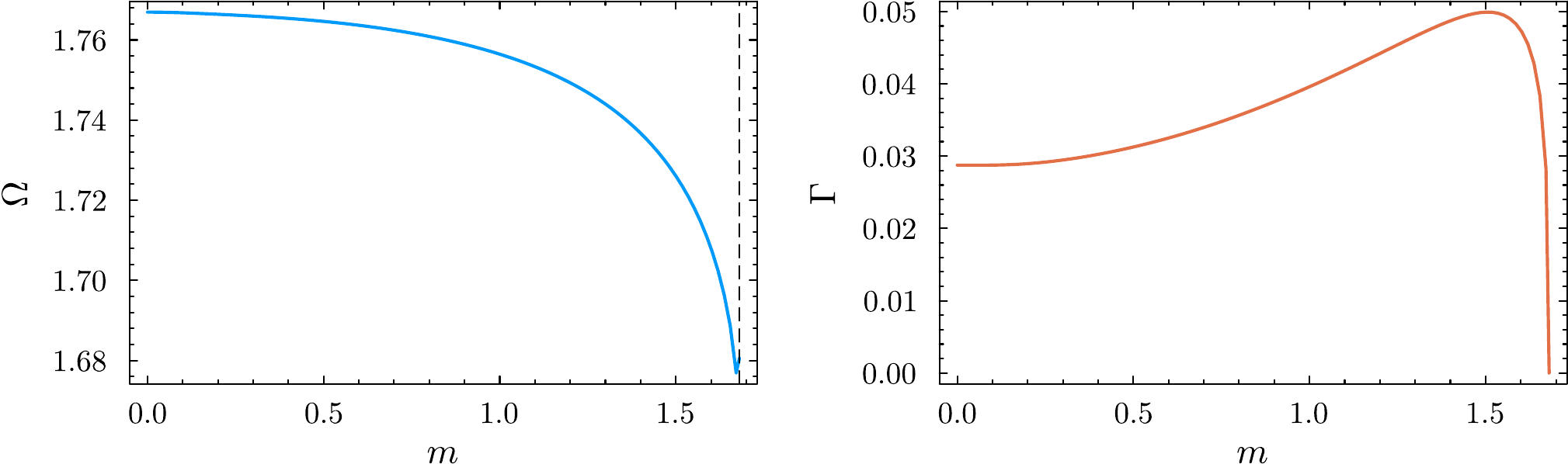}
    \caption{Spectral structure of the lowest QNM with the frequency defined as $\omega=\Omega+i\Gamma$ for  $\varepsilon=0.01$ in the rational model.}
    \label{fig:rational_spectralQNM_mass}
\end{figure}

Figures~\ref{fig:rational_spectral}~and~\ref{fig:rational_spectralQNM_mass} show the real and imaginary parts of the lowest QNM as a function of the parameter $\varepsilon$ with $m = 0$ fixed, and as a function of the parameter $m$ with $\varepsilon = 0.01$ fixed, respectively.

\section{Polynomial $\phi^{10}$ model}
\label{polphi10}
\begin{figure}
    \centering
    \includegraphics[width=1\linewidth]{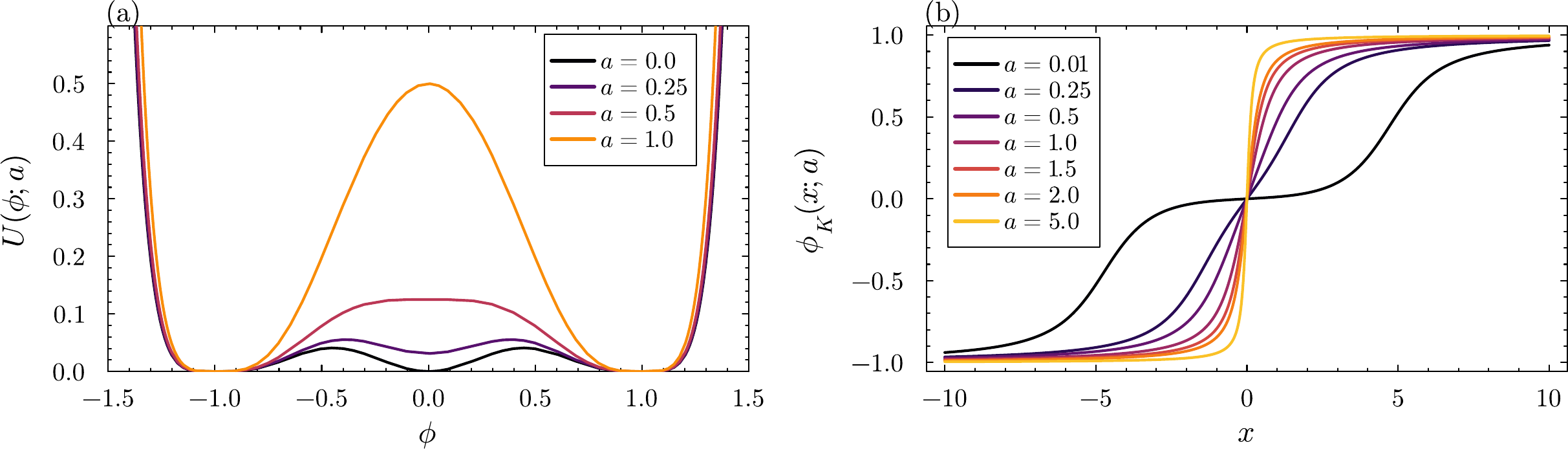}
   
    \caption{(a) Potential as a function of $\phi$ and (b) kink configuration for several values of $a$.}
    \label{fig:phi10_pots}
\end{figure}

Let us also consider a $\phi^{10}$ polynomial model, as a more natural model compared with rational models. The potential is given by
\begin{equation} \label{phi10pot}
    U(\phi;a)=\frac{1}{2}(\phi^2-1)^4(\phi^2+a^2).
\end{equation}
It has two minima, at $\phi=\pm 1$. For $a=0$, the potential acquires an extra minimum at $\phi=0$. In Fig.~\ref{fig:phi10_pots}(a), we show the potential as a function of the field $\phi$ for several values of the parameter $a$, and in Fig.~\ref{fig:phi10_pots}(b), we present the possible kink solutions. 
For $a>\frac{1}{2}$ the potential has two equal minima at $\phi=\pm 1$. For $a\in(0, \frac{1}{2})$ the potential has a local minimum at $\phi=0$ which is a false vacuum with mass $m_f=\sqrt{1-4a^2}$. For $a=0$, the potential has three equal minima, similar to the standard $\phi^6$ model, but two of the vacua are massless and one is massive. For $a<0$, the vacuum at $\phi=0$ is the only true vacuum of the model, and no topological defects can exist.
We will mainly focus on $a>0$ here.

It is well-known that a local minimum in the potential leads to an internal structure, as can be seen in the kink profiles. When the value of the potential at the local minimum is close to zero, the kink may be viewed as a composite object made of two subkinks. Other examples in the literature where the same phenomenon occurs include the double sine-Gordon \cite{campbell1986kink} and Christ-Lee \cite{dorey2021resonance} models.

Solving the BPS equation given by
\begin{equation}
    \frac{d\phi}{dx} = \sqrt{\phi^2+a^2}(1-\phi^2)^2,
\end{equation}
for the \ref{phi10pot} potential, gives the following solution
\begin{align}
x - A \;=\; \frac{a^{2} + 2}{4 (a^{2} + 1)^{3/2}}
\, \ln\!\left[
\,\frac{\,1 + \sqrt{1 + a^{2}} - \phi^2+\phi \sqrt{a^{2} + \phi^{2}}}
{1 - \sqrt{1 + a^{2}} - \phi^{2} +\phi \sqrt{a^{2} + \phi^{2}}}
\right]
\;-\;
\frac{\phi \, \sqrt{a^{2} + \phi^{2}}}{2 (a^{2} + 1)(\phi^{2} - 1)} \, .
\end{align}
From the above expression, we can find the value of $x$ for any arbitrary $-1<\phi<1$.
Let us define once more $\phi_K(x)=1-F(x,a)$. The expansion around the tail at large positive $x$, taking $F(x,a) \ll 1$ gives
\begin{align}\label{tail-phi10}
F(x,a) \;=\; \frac{1}{4\sqrt{1+a^{2}}\,(x-A)}
\;+\; \frac{a^{2}+2}{16(1+a^{2})^{2}} \,
\frac{\ln(\Lambda(x-A))}{(x-A)^{2}}
\;+\; \mathcal{O}\left((x-A)^{-2}\right),
\end{align}
where 
\begin{align}
\Lambda\;=\; \frac{8\,(1+a^{2})^{3/2}}{\big(\sqrt{\,1+a^{2}\,}-1\big)^{2}}.\nonumber
\end{align}

\begin{figure}
    \centering    \includegraphics[width=0.6\linewidth]{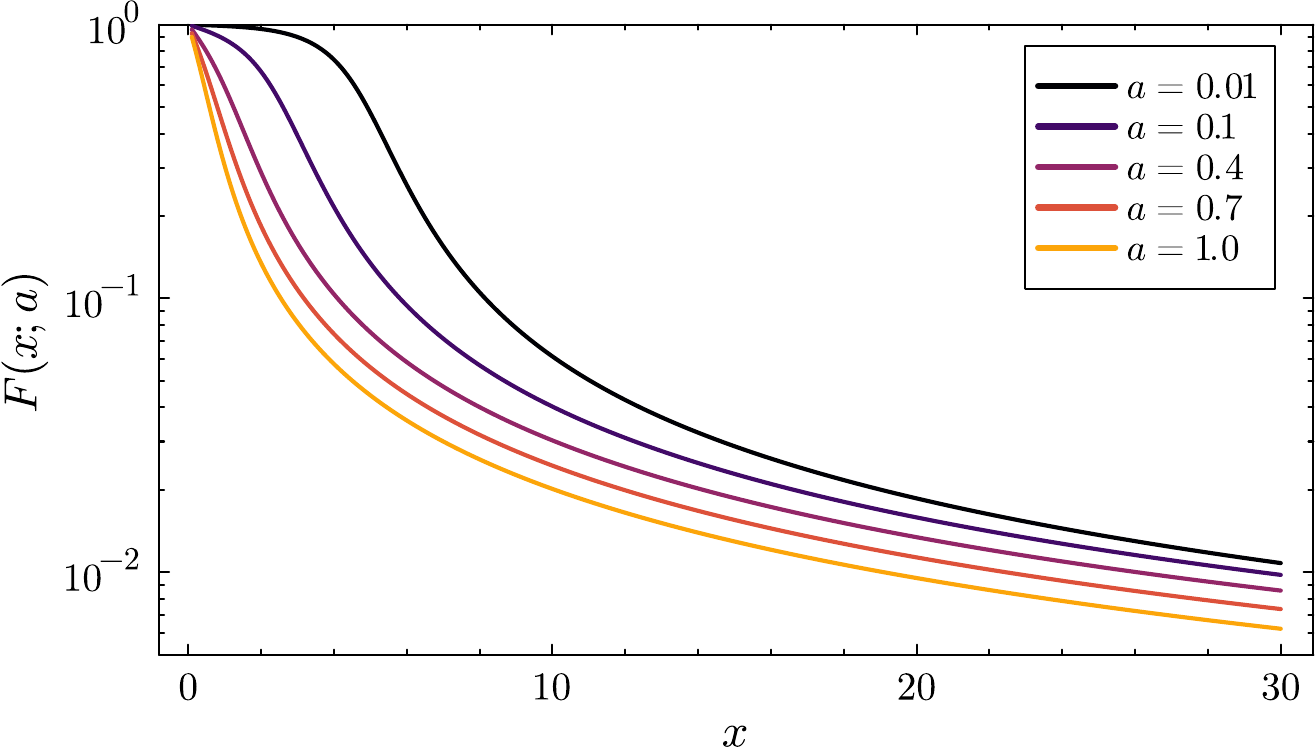}
    \caption{Kink tail as a function of $x$ for large positive $x$, for several values of $a$ in $\phi^{10}$ model.}
    \label{fig:phi10-tail}
\end{figure}

Figure~\ref{fig:phi10-tail} shows this tail behavior on a logarithmic scale for several values of $a$. The expression above clearly illustrates the long-range nature of the tail.

\begin{figure}
    \centering
    \includegraphics[width=0.665\linewidth]{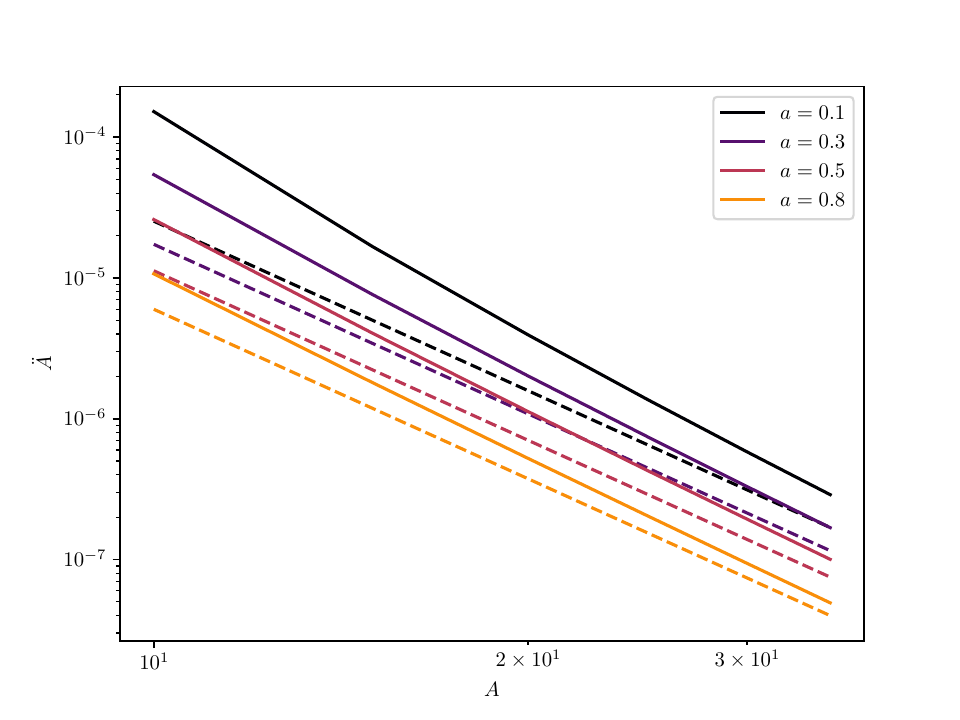}
   
    \caption{Acceleration as a function of $A$ for several values of $a$. The solid lines are the numerical results from the kink-antikink dynamics and the dashed lines are theoretical approximation in \ref{accel-manton}.}
    \label{fig:phi10_accel}
\end{figure}

We can estimate the force between the kink and antikink, using the Manton method in \cite{manton2019forces}, looking only at the kink tail facing the antikink. One can approximate the kink by a propagating field $\eta(y) \equiv \phi(x-v t)$ in this method. Taking only the first term in \ref{tail-phi10}, we can approximate the acceleration in the following form
\begin{align}\label{accel-manton}
\ddot A=-\frac{1}{32(1+a^2)E_{BPS}}\left(\frac{\Gamma[1/4]^2}{\sqrt{32 \pi}}\right)^4 A^{-4}.
\end{align}
The BPS energy in the above expression is given by
\begin{align}
E_{BPS}&=W(1)-W(-1)\nonumber\\
&=\frac{1}{24} \left[ \sqrt{1 + a^2} \left(8 - 10a^2 - 3a^4\right) + 3a^2 \left(8 + 4a^2 + a^4 \right) \, \mathrm{arcsinh}\left(\frac{1}{a}\right)\right],
\end{align}
defining $U(\phi;a)=W_\phi^2/2$ where the subscript represents the derivative with respect to $\phi$. The force simply is $F_{\phi^{10}}=E_{BPS}\ddot A$, which scales as the inverse fourth power of the kink position, as expected \cite{gonzalez1989kink, mello1998topological, manton2019forces}. In Fig.~\ref{fig:phi10_accel}, we compare the theoretical prediction for the acceleration from \ref{accel-manton} with the numerical results obtained from kink–antikink dynamics, shown on a logarithmic scale. As can be seen, the agreement is better at large separations, which is natural since, at smaller separations, the second term in the expansion \ref{tail-phi10} becomes important.  

\subsection{Spectral Analysis}
\label{sec:specphi10}
\begin{figure}
    \centering
    \includegraphics[width=0.5\linewidth]{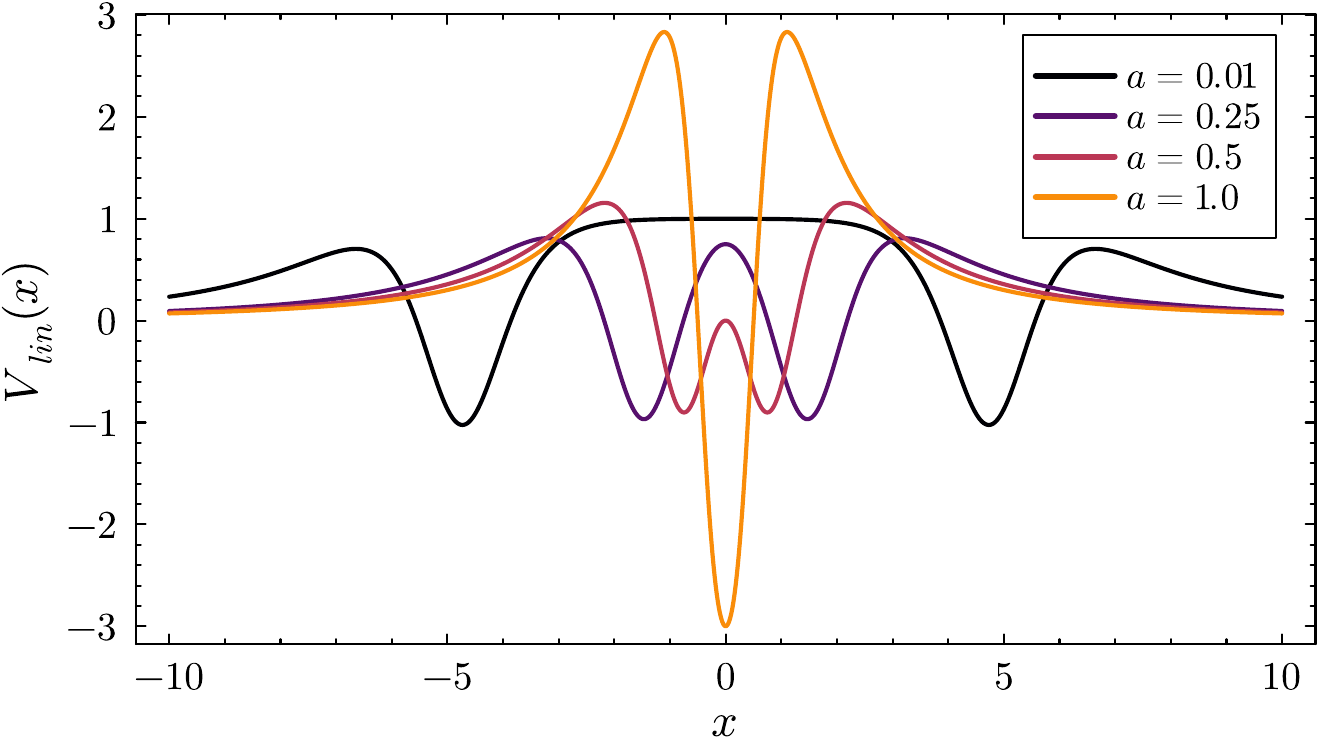}
    
    \caption{Stability potential for several values of $a$ in $\phi^{10}$ model.}
    \label{fig:phi10_effpots}
\end{figure}

The $\phi^{10}$ model considered here has a poorer spectral structure when $a\neq 0$ , as it only supports long-range kinks on both tails. However, as we are looking for resonance via QNMs, it perfectly serves our purpose. Figure~\ref{fig:phi10_effpots} shows the linear stability potential as a function of $x$ varying the parameter $a$. As one can see, the potential does not admit any bound states. Apart from the continuum of scattering modes, only the possibility of QNMs exists.  

\begin{figure}
    \centering
    \includegraphics[width=1\linewidth]{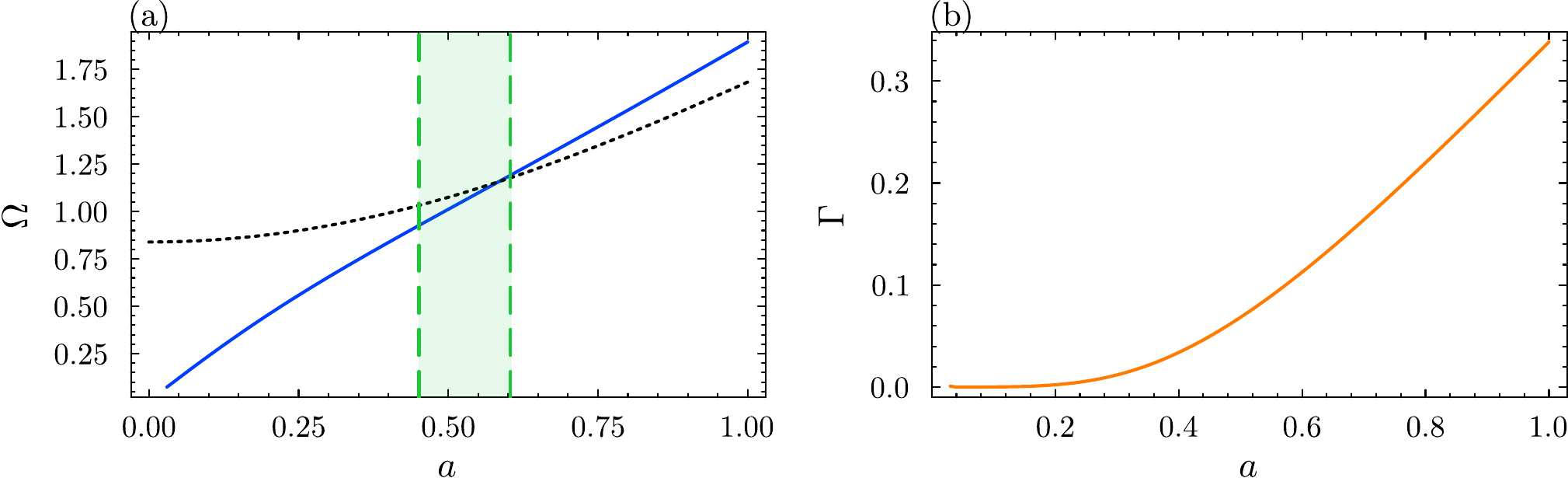}
    
    \caption{Lowest QNM in $\phi^{10}$ model. The left panel shows the real part, and the right panel shows the imaginary part of the frequency.}
    \label{fig:phi10_qnms}
\end{figure}

Figure~\ref{fig:phi10_qnms} shows the real and imaginary parts of the lowest QNM. In the left panel, the dotted curve shows the maximum of the stability potential as a function of the parameter $a$. The curve of the real part below this threshold may be interpreted as the frequency of the mode bound to the kinks. The green strip will be important for the result of the collision.

Interestingly, when $a$ is small, a clear relationship emerges between the lowest QNM and the subkink structure. The QNM corresponds to the antisymmetric combination of the subkink's zero modes. As expected for a Schrödinger equation with double-well potentials, the two lowest modes are the symmetric and antisymmetric combinations of the single-well ground state. The symmetric combination corresponds to the true zero mode of the compound kink, and the antisymmetric combination corresponds to its lowest QNM. Therefore, as the internal structure becomes more pronounced and the subkinks fully separate, the QNM should turn into a zero mode. This implies that both its frequency and decay rate must vanish in this limit, which agrees perfectly with the results in Fig.~\ref{fig:phi10_qnms}.

\section{Kink-Antikink Collisions}
\label{sec:res}

We aim to investigate whether resonance windows can emerge in kink-antikink collisions involving kinks with long-range tails on both sides. We are also interested in whether QNMs can create resonant behavior in this more extreme scenario. To this end, we will study collisions in two families of models introduced in the previous sections: the family of rational models, and subsequently the $\phi^{10}$ model.

\subsection{Rational Model}

\begin{figure}
    \centering
    \includegraphics[width=0.63\linewidth]{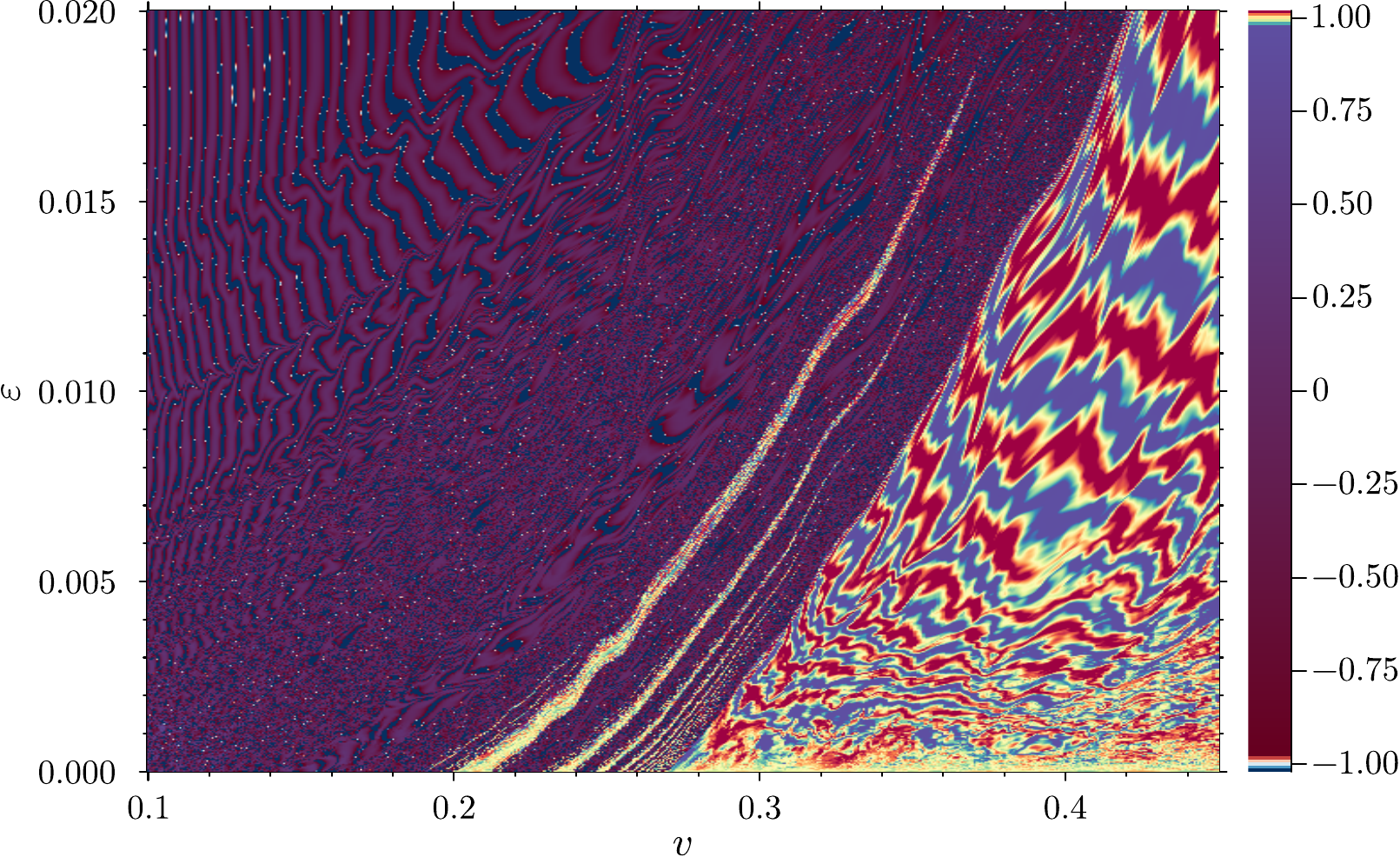}
    \caption{Field at the center of mass and final collision time as a function of $v$ and $\varepsilon$. We fix $m=0$.}
    \label{fig:phi-vs-e2}
\end{figure}

\begin{figure}
    \centering
     \includegraphics[width=0.6\linewidth]{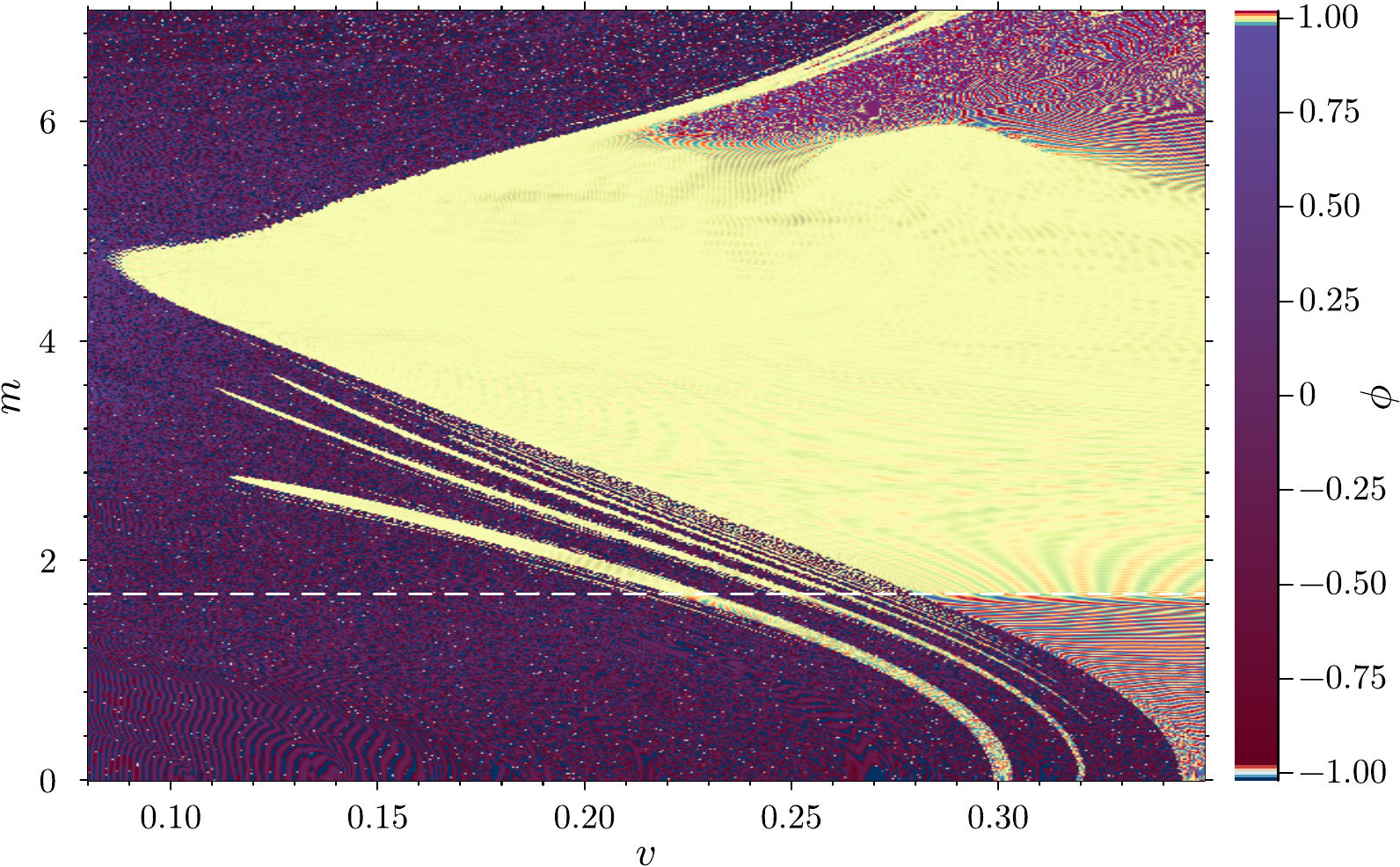}
    \caption{Field at the center of mass and final collision time as a function of $v$ and $m$. We fix $\varepsilon=0.01$. The white dashed line indicates the value of $m$, $m_w=1.69776$ at which the BM vanishes at the threshold.}
    \label{fig:phi-vs-m2-pal}
\end{figure}

We identify the scattering output via the field value $\phi(0,t_f)$, i.e., at the center of mass $x=0$ and at the final collision time $t_f$. The parameter space is very large, as it involves two free parameters, $m$ and $\varepsilon$.  To systematically explore this space, we first fix $m$ and vary $\varepsilon$, and then fix $\varepsilon$ and vary $m$. The corresponding results are presented in Figs.~\ref{fig:phi-vs-e2} and \ref{fig:phi-vs-m2-pal}, respectively.


\begin{figure}
    \centering
    \includegraphics[width=0.6\linewidth]{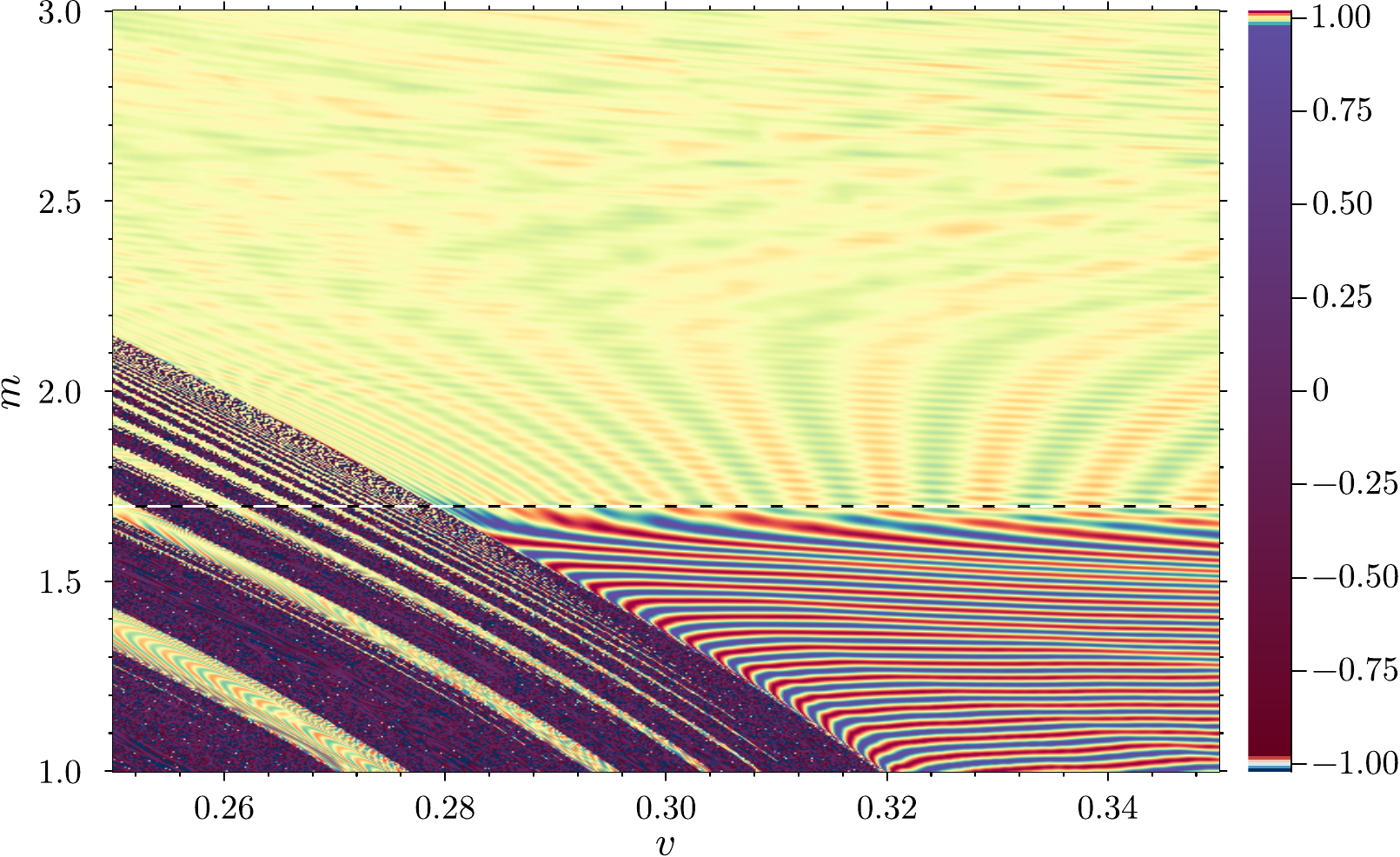}
    \caption{Contour plot around $\phi$ = 1 at the center. It shows the wave patterns in Fig.~\ref{fig:phi-vs-m2-pal} around the line where the BM turns into the lowest QNM. Observe how the wave pattern changes around this line.}
    \label{fig:phi-vs-m2}
\end{figure}



The kinks are only weakly long-range for this model for small $\epsilon$. Hence, the initial conditions are obtained from the simple split domain ansatz, without any minimization. They read
\begin{align}
    \phi(x,t)=H(-x)\phi_K(\gamma(x+x_0-vt))-H(x)\phi_K(\gamma(x-x_0+vt)),
\end{align}
where $H(x)$ is the Heaviside function. The half inter-kink separation is $x_0=25$, and we fix the final collision time at $t_f=80/v$.

In Fig.~\ref{fig:phi-vs-e2}, we fix $m=0$ and choose $\varepsilon\in[0,0.02]$. In this case, the kinks are long-range on both sides for any finite $\varepsilon$. However, the long-range character occurs in a limited region for small $\varepsilon$, gradually increasing for larger $\varepsilon$. Consequently, we observe resonance windows on the short interval $0\leq\varepsilon\lesssim 0.018$. For higher values of $\varepsilon$, the resonance windows are suppressed.

In Fig.~\ref{fig:phi-vs-m2-pal}, we fix $\varepsilon=0.01$ and choose the mass gap $m\in[0,7]$. In this scenario, the kinks have short-range tails, except for the zero mass gap case ($m=0$). As $m$ decreases all the way to $m=0$, some resonance windows are suppressed, but some windows survive up to the point where the tails become long range. 

Therefore, we have shown that this family of rational models does exhibit resonance windows, despite the kinks being long-range on both sides and having a zero mass gap. The windows are only possible because of the existence of QNMs.  

\subsection{Polynomial $\phi^{10}$ model}

For the polynomial $\phi^{10}$ model, the initial conditions for the collision were obtained from the convection-diffusion algorithm, a novel method which we describe in Section \ref{sec:conv-diff}. To ensure the precision of the method, we have compared the numerical output with a previously established method in the literature, the one with two layers of minimization \cite{christov2019long, christov2019kink, christov2021kink, campos2021interaction}. The scattering outputs from both algorithms are virtually indistinguishable. A more quantitative comparison between the methods is provided in the subsequent section.

\begin{figure}
    \centering
    \includegraphics[width=0.6\linewidth]{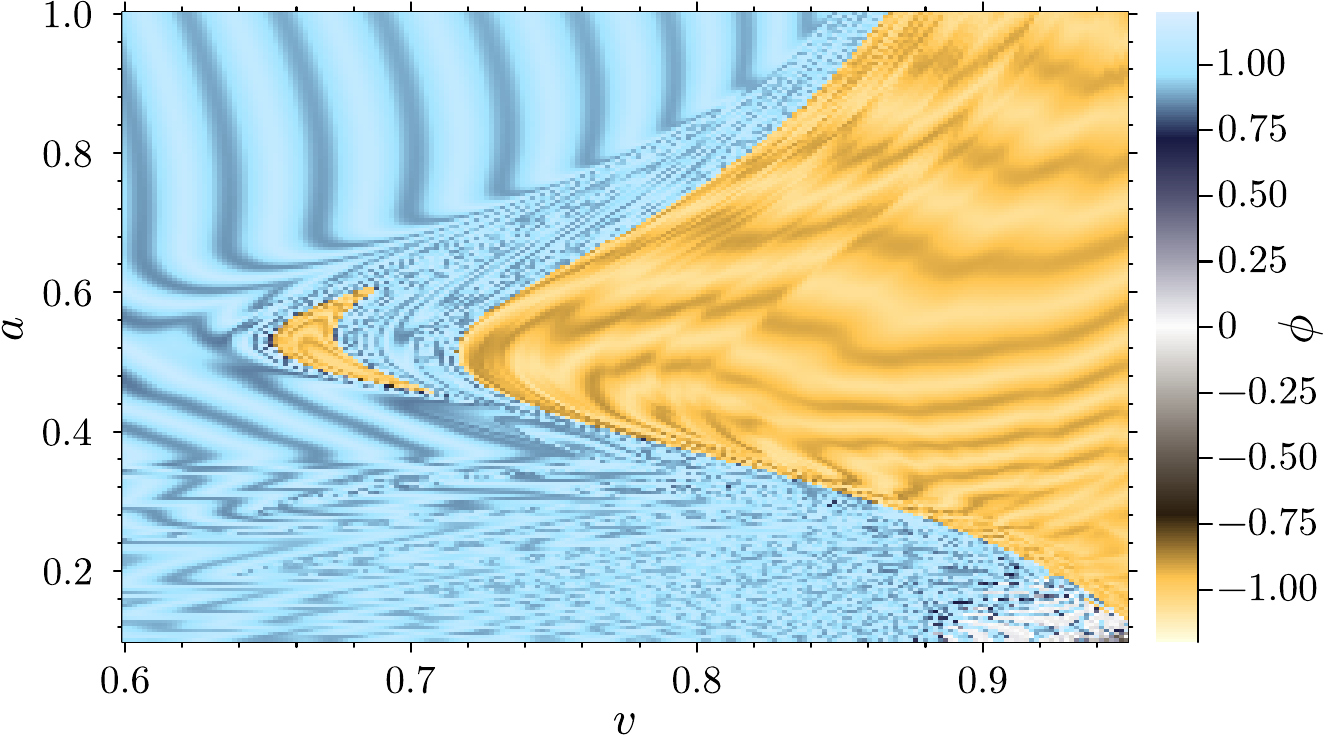}
    \caption{Field at the center of mass and final collision time as a function of $v$ and $a$. The potential is $\phi^{10}$ polynomial.}
    \label{fig:phi10}
\end{figure}

The field at the center of mass and final time $\phi(0,t_f)$ is shown as a function of $v$ and $a$ in Fig.~\ref{fig:phi10}. The blue region marks the annihilation with short-lived bions and a large amount of radiation. We will show that there is no bion formation for large values of $a$, and kink-antikink turns directly into radiation after the collision. This behavior is characteristic of kink collisions with strongly long-range tails on both sides \cite{campos2021interaction}. The yellow color, on the other hand, describes kink separation. In the larger yellow region, the separation occurs after a single bounce. Interestingly, the output also shows an isolated yellow island that exhibits two-bounce resonance windows. The island is centered around $(a,v)=(0.55, 0.67)$.

The resonant exchange mechanism occurs through the excitation of a QNM, which has been located in the linear stability section. Therefore, resonance windows do occur in polynomial models with long-range tails on both sides through the excitation of QNMs. This provides another instance of models with zero mass gap that exhibit resonance windows, but this time the tails are strongly long-range. For this reason, the resonance windows are more fragile, so there is only a single window that exists only in a small region of the parameter space.

The result in Fig.~\ref{fig:phi10} can be better understood by analyzing the lowest QNM shown in Fig.~\ref{fig:phi10_qnms}. For $a>0.6$, the real part of the QNM frequency exceeds the maximum stability potential, hence unbinding from the kinks. In this regime, the internal structure is absent, and both the QNM frequency and decay rate are too large to allow the formation of resonance windows. For small values of $a$, resonance windows are also suppressed, though for different reasons. Two main factors may contribute. First, the real part of the frequency becomes small, indicating that the energy transfer mechanism may need a higher amplitude, giving rise to more pronounced anharmonic oscillations. Besides that, the internal structure plays a more important role as the subkinks start becoming individual identities. This may mean that the kink-antikink collision for small $a$ behaves closer to the dynamics of two pairs of kink-antikink. In this case, the system dynamics becomes more complex \cite{dorey2023collisions} and hinders the appearance of resonance windows.

\begin{figure}
    \centering
    \includegraphics[width=1\linewidth]{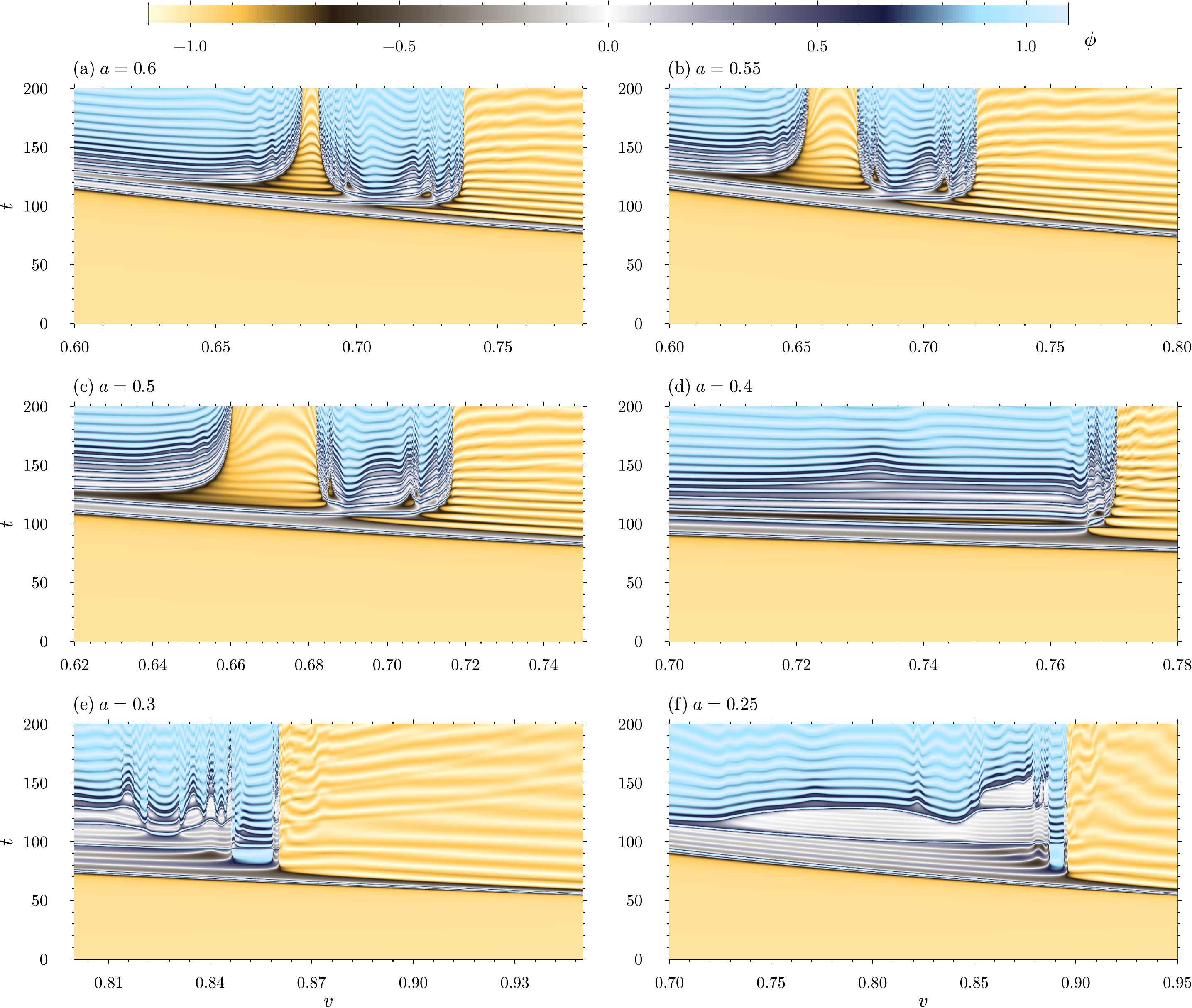}
    \caption{The field at the center of mass as a function of time and initial velocity.
    }
    \label{fig:phi-vs-tv2}
\end{figure}

Fig.~\ref{fig:phi-vs-tv2} shows the center of mass dynamics of the field as a function of the initial velocity for several values of the parameter $a$. In the region with the single resonance window, we were able to locate a few false resonance windows in the scattering output to estimate the resonant frequency $\omega_r$. Each bounce is represented in the figure by a light blue line. If you start counting from the bottom of the figure, the third-bounce is represented by the third blue line. A false two-bounce window can be located by a bump in that line, representing that the kinks took a longer time to bounce a third time, that is, the kinks nearly fully detached. 

So we considered two sets of resonance windows, both true and false, for $a=0.5$ and $a=0.55$, obtaining the frequencies $\omega_r=0.992\pm0.003$ and $\omega_r=1.061\pm0.093$, respectively. The error can be quite large because very few windows were found. The result should be compared with the theoretical QNM frequencies obtained in the linear stability analysis. They are $\omega_{th}=1.01156$ and $\omega_{th}=1.09810$, respectively, being remarkably close to the numerical values. It is astonishing that the resonant energy exchange mechanism works so well in this extreme scenario.


\begin{figure}
    \centering
    \includegraphics[width=\linewidth]{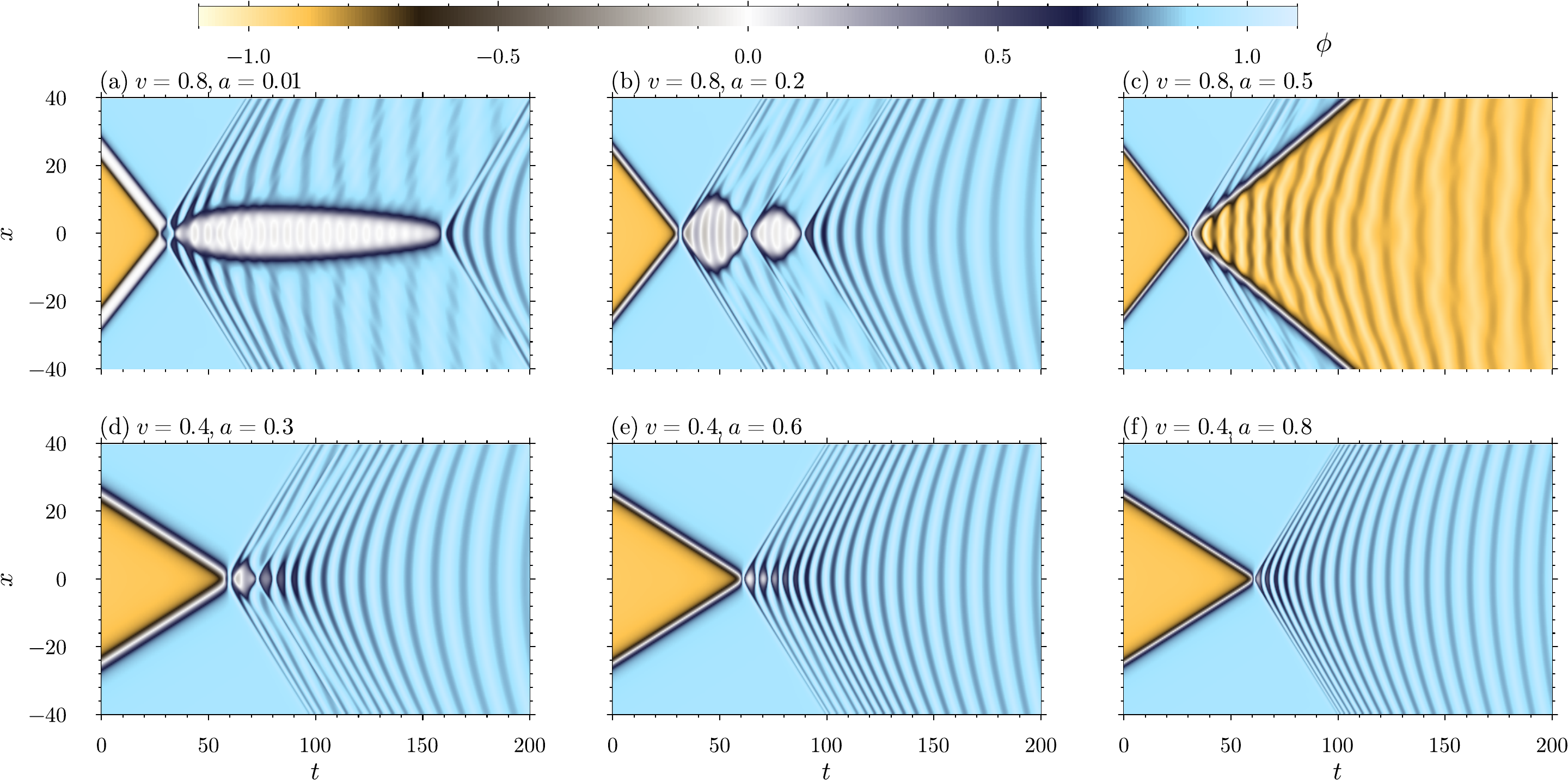}
    \caption{Field dynamics for several values of $a$ and initial velocity.}
    \label{fig:bion}
\end{figure}

In Fig.~\ref{fig:bion}, we present the field dynamics for different values of $a$ and initial velocities. The figure shows that, in addition to the resonance window being fragile, bion formation in kink–antikink collisions is also fragile when the kinks have long-range tails on both sides. In the class of models studied in \cite{campos2021interaction}, the kinks annihilate directly into radiation without forming bions, except for the appearance of one bion turning into radiation at the second bounce for ultrarelativistic initial velocity. This is mainly due to the long-range character of kinks' backtails. Here,  Fig.~\ref{fig:bion} shows that bion formation is possible, but it is very short-lived whenever it occurs. After only a few bounces, it decays into radiation. No bion is formed for large $a$, and kinks annihilate directly into radiation. For small $a$, as the kink acquires internal structure and the subkinks become closer to individual identities, bion formation becomes increasingly rare and requires high initial velocities to occur.

\section{Convection-diffusion Algorithm}
\label{sec:conv-diff}

One of the standard methods for finding stationary point solutions through energy minimization is the gradient flow method. For a generic energy functional $E[\phi]$, the minimization follows the following equation
\begin{equation}
\alpha\frac{\partial \phi}{\partial t} = -\frac{\delta E}{\delta \phi}.
\end{equation}
In the case of a scalar field theory with a potential $V(\phi)$, this equation takes the form
\begin{equation}
    E[\phi]=\int_{-\infty}^\infty \left[\frac{1}{2}\phi_x^2 + V(\phi)\right]\,dx,
\end{equation}
for static configurations.
The gradient flow equation becomes a nonlinear diffusion equation where the right-hand side is just a static equation of motion
\begin{equation}
    \alpha\phi_t  = \phi_{xx}-V'(\phi).
\end{equation}
The gradient flow method evolves some generic initial conditions and relaxes the configuration until the static equations of motion are satisfied.
\begin{figure}
    \centering
    \includegraphics[width=1\linewidth]{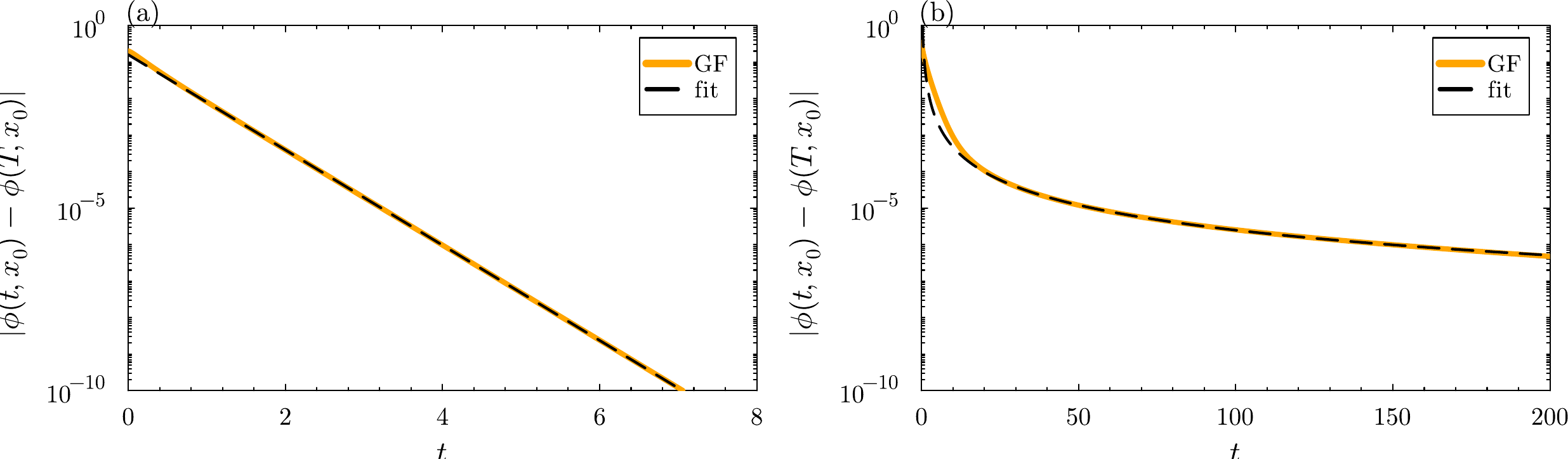}
    
    \caption{Gradient flow convergence test for the (a) $\phi^4$ model ($T=20$) and $\phi^{10}$ model ($T=400$). In both cases $x_0=1$. }
    \label{fig:gradient}
\end{figure}

We have tested the gradient flow (GF) method, shown in Fig.~\ref{fig:gradient}, to check the convergence for $\phi^4$ and $\phi^{10}$ models and found kink solutions starting from the initial guess $\phi_{init}(x)=\tanh(2x)$. In the case of the standard massive $\phi^4$ case, we found that the GF method converges exponentially for fixed $x_0=1$. The best fit gives
\begin{equation}
    \phi_{fit}(x_0,t) = \phi_{final}+0.1601 \exp(-3.0048t),
\end{equation}
which is consistent with the fact that in the case of existing bound modes, the GF method should converge as $\sim e^{-\omega^2t}$ and the kink in the $\phi^4$ model has an internal mode with $\omega=\sqrt{3}$.
In the case of the massless $\phi^{10}$ model, the convergence is much slower. It is no longer exponential. The best power law fit gives
\begin{equation}
    \phi_{fit}(x_0,t) = \phi_{final} + 0.0918 t^{-2.2837}.
\end{equation}
This shows that even static solitons in massless models are more difficult to find using the gradient flow method.

In our case, we need to find a solution that is not static but rather moving with a constant velocity $v$. In this case, we need to replace the spatial derivative with $\partial_x\to\gamma^{-1}\partial_x$, where $\gamma$ is the Lorentz factor. The gradient flow equation takes the form
\begin{equation}
    \alpha\phi_t  = (1-v^2)\phi_{xx}-V'(\phi).
\end{equation}
The additional factor $(1-v^2)$ in front of the second derivative is responsible for Lorentz contraction.  
The above equation has three stationary solutions depending on initial conditions: a vacuum (in case of initial conditions with charge zero) or a kink/antikink (in case of initial conditions with charge one/minus one).
In order to describe colliding kink and antikink (charge zero), we assume even symmetry of the field $\phi(-x)=\phi(x)$ and restrict to a half-line $x>0$. One can take any configuration resembling a widely separated kink-antikink pair for the initial conditions. In such a case, the diffusion-like equation acts in two ways: it smoothens out the fluctuations of the field around the defects, and it also attracts the kink and antikink towards each other. The latter is due to the fact that kink and antikink attract each other through a force that is small in distance between them. Therefore, if the initial configuration is not exactly a kink-antikink pair, the diffusion-like equation will make them approach each other. We can assume that this drag is slow compared to the relaxation of the fluctuations, so that the ultimate annihilation of the pair takes place at very large times. 
However, we would like to have more control over the position of the defects. Waiting for a very long time until they move to their position is also not very practical. Therefore, we introduce an additional convection term into the equation as follows
\begin{equation}
    \alpha\left(\phi_t  - v\phi_x\right) = (1-v^2)\phi_{xx}-V'(\phi),
\end{equation}
which forces the solution to move with velocity $-v$. By solving this equation, we obtain a configuration describing a kink-antikink pair moving towards each other with velocity $v$
with minimal fluctuations around them and in the center. The arbitrary parameter $\alpha$ can be used to tune the speed of the relaxation process. The smaller the parameter, the faster the relaxation. However, values that are too small can lead to numerical instabilities. With trial and error, we have found that $\alpha=1$ is a good choice.

Note that the additional force $F_{\phi^{10}}$ between the kink and antikink changes the velocity by
\begin{equation}
   \alpha \Delta v = F_{\phi^{10}}
\end{equation}
This value can be neglected because it is approximately $7\cdot10^{-2}A^{-4}$ for values of $a\approx 0.5$. $A$ can be treated as a position of the kink, which, for our initial conditions was $A=25$.

A quantitative comparison between the Convection-Diffusion and other established methods in the literature for long-range kinks' collisions was performed as follows. First, we perform the tests in the following $\phi^{10}$ model
\begin{equation}
    \mathcal{L}=\frac{1}{2}\partial_\mu\phi\partial^\mu\phi-\frac{1}{2}(\phi^2-1)^5,
\end{equation}
which has been previously considered in Ref.~\cite{campos2021interaction}. In this model, the kinks are also long-range on both sides. The $\phi^{10}$ model above was chosen due to its slower tail decay, as will be clarified below.

We now report on the results of the comparison of three different methods. We will briefly discuss the concepts involved and refer the reader to Refs.~\cite{christov2019kink,christov2019long,christov2021kink,campos2021interaction,campos2024collisions} for further details. The employed methods are the split-domain (SD) ansatz, Convection-Diffusion (CD), and two layers of minimization developed in Refs. \cite{christov2021kink, campos2021interaction}. The SD ansatz is the best nonoptimized method to simulate long-range kink collisions \cite{christov2019long}.

The first comparison test is as follows. We measure how well the solutions obey the Lorentz contracted static equation via the following Euclidean norm
\begin{equation}
    F[\phi]=\lVert(1-v^2)\frac{d^2\phi}{dx^2}-V'(\phi)\rVert^2_2,
\end{equation}
 at half-separation $x_0=50$, as a measure of the error in the initial field. The norm is found by considering the field as a vector and computing the derivatives by finite differences. The CD algorithm decrease the norm by a factor of $\mathcal{O}(10^2)$, while the two layers of minimization decreases the norm by a factor of $\mathcal{O}(10^8)$. The smaller accuracy occurs because the kink is moving in the CD algorithm. Thus, it passes quickly by each postion. As we are looking for the optimal configuration at a fixed position, the CD algorithm performs this task slightly sub-optimally. Then, we measure how well the solutions obey the Lorentz contracted zero-mode equation
\begin{equation}
    G[\chi;\phi]=\lVert(1-v^2)\frac{d^2\chi}{dx^2}-V''(\phi)\chi\rVert^2_2,
\end{equation}
as a measure of the error in the initial velocity field $\chi=\dot\phi$.
The Convection-Diffusion decreases the norm by a factor of $\mathcal{O}(10^4)$ with respect to the SD ansatz, which is a larger factor than the one from the two layers of minimization $\mathcal{O}(10^3)$. 

\begin{figure}
    \centering
    \includegraphics[width=0.75\linewidth]{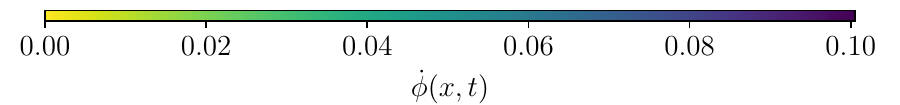}
    \includegraphics[width=\linewidth]{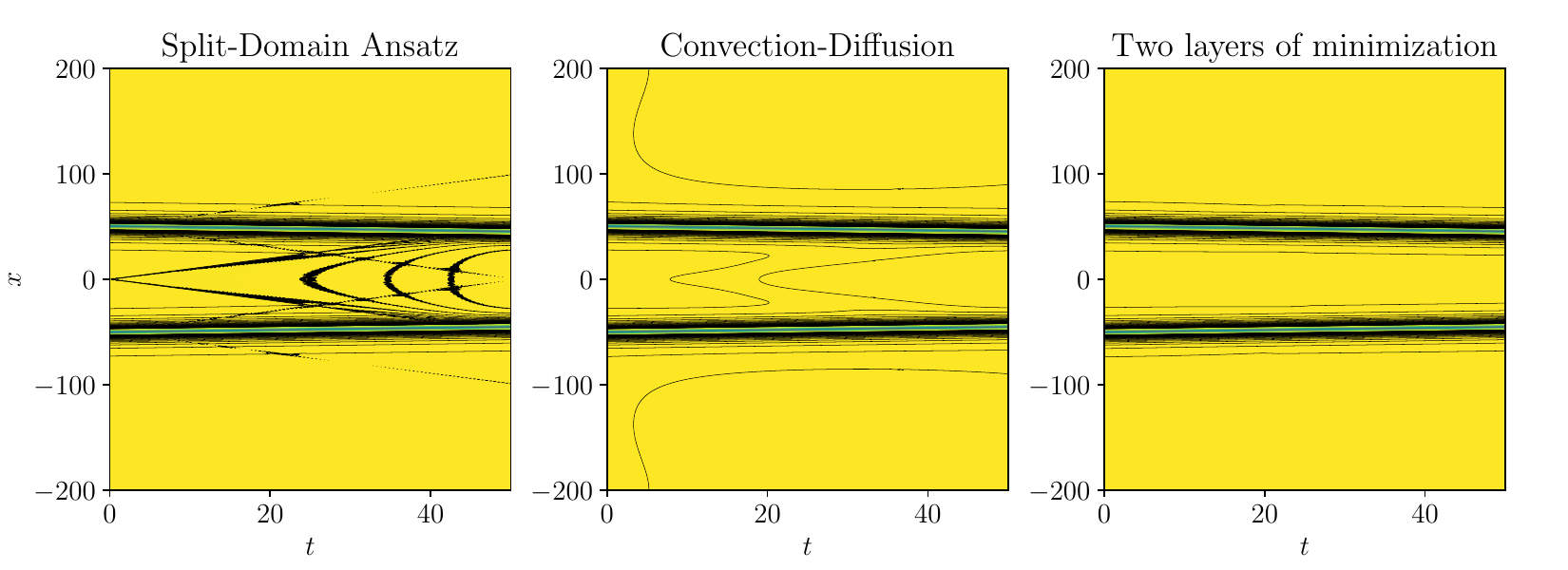}
    \caption{Velocity field $\dot\phi(x,t)$ evolution in spacetime. The kinks are initialized according to the SD ansatz, Convection-Diffusion algorithm, and two layers of minimization. We also draw 41 contour lines uniformly separated in the interval $[0.0,0.004]$.}
    \label{fig:comparison}
\end{figure}

The second comparison test consists of integrating the equations of motion in time and comparing the actual field evolution. As radiation is more easily spotted on the velocity field $\dot\phi(x,t)$, we show its colormaps and contour plots in Fig.~\ref{fig:comparison}. We draw 41 contour lines uniformly separated in the interval $[0.0,0.004]$. The SD ansatz generates a large amount of radiation, whereas the CD contains radiation only at the innermost and outermost contours. The method with two layers of minimization serves as a reference, showing no radiation at this scale. It is important to emphasize that the contour plots show a very fine structure of the field evolution and, if we were simulating less long-range kinks, it would have been even more difficult to locate radiation in the CD simulations. So we conclude that the CD method has a high degree of accuracy.

Although the CD method is less precise than the two-layer minimization, it can be advantageous for two reasons. First, and most importantly, the minimization is usually performed through numerical algorithms, which often do not have a clear physical interpretation. The Convection-Diffusion algorithm, in contrast, does have a clear physical interpretation. The first-order dynamics provides damping that absorbs most radiation, while leaving the propagating wave behavior with velocity $v$ undamped by construction. Second, the Convection-Diffusion is a time-efficient method, because it is an initial value problem for a partial differential equation that is first-order in time. Therefore, it is always more time efficient than the actual time integration with second-order dynamics and thus will never be a bottleneck for large-scale simulations.

\section{Conclusion}
\label{sec:conc}

In this paper, we studied a scalar field theory in $(1+1)$-dimensions, with two distinct scalar potentials, both with the possibility of admitting kinks with long-range tails on both sides. First, we considered a rational model where the potential has two symmetric minima and two control parameters $m$ and $\varepsilon$. We showed that the kink solutions of this model may exhibit either short-range or long-range tails, depending on the values of these parameters. In particular, when $m=0$ and $\varepsilon>0$, the vacua become massless, and the corresponding kinks become long-range on both sides; however, in a finite spatial interval, whose size increases with $\varepsilon$. As a more realistic example of long-range kinks with fatter tails, we also considered a $\phi^{10}$ model with a single free parameter $a$. We analyzed the spectral structure of small perturbations around the kink for both models. The rational model exhibits a significantly more complex spectrum than the $\phi^{10}$ model. Depending on the values of $m$ and $\varepsilon$, it can support bound, antibound, and quasinormal modes.
However, both models admit only QNMs for kinks with long-range tails on both sides. 
We have shown that there are resonance windows in the kink-antikink collision dynamics, where the lowest QNM is responsible for the energy exchange mechanism. In the rational model with $m=0$, these windows appear only within a narrow range of $\varepsilon$ just above zero. As $\varepsilon$ increases, the kinks become increasingly long-range, and the resonance windows disappear. In the $\phi^{10}$ model, a single resonance window emerges within a relatively short interval of $a$, centered around $a \approx 0.55$. Above this interval, no resonance window is observed, as the real part of the lowest QNM frequency exceeds the maximum value of the linear stability potential. The disappearance of the windows below this interval is mainly due to the kinks' internal structure, which appears as subkinks. For small values of $a$, these subkinks behave as nearly independent entities, leading to more complex dynamics. 
By taking not only the true resonance window but also a few existing false windows, we compared the frequencies governing the energy transfer mechanism with the real part of the lowest QNM frequency. The agreement turned out to be better than expected, considering the few points from both the false and true windows.

Interestingly, we found that the wave pattern changes in the kink-antikink dynamics in the $\phi^{10}$ model, when the lowest normal mode turns into the QNM. As a consistency check, we also compared the kink–antikink acceleration obtained numerically from the dynamics with the theoretical prediction based on the Manton method. The two results show good agreement at large separations, as expected. Besides that, we looked at the kink-antikink annihilation with or without the bion formation in different $a$ regions. Our analysis shows that both resonance windows and bion formation in kink–antikink collisions are fragile when the kinks possess long-range tails on both sides. While bions can occasionally form, they are short-lived, decaying into radiation after only a few bounces. For large $a$, no bions appear and the kinks annihilate directly into radiation, whereas for small $a$, the emergence of subkinks makes bion formation increasingly rare and restricted to high initial velocities.

Here, we also proposed a new algorithm for initializing long-range kink collisions, based on convection–diffusion dynamics. Compared to the split-domain ansatz, this method produces higher-quality initial data by suppressing unphysical radiation while keeping computational efficiency. Although it is less accurate than the two-layer minimization technique, the convection–diffusion approach has the advantage of a clear physical interpretation and substantially lower computational cost, making it a suitable method for large-scale simulations of long-range soliton dynamics.

There remain many important questions to explore in future works. For instance, how bion formation relates to the energy exchange mechanism and resonance windows. Another path worth taking could be to better understand this mechanism via QNMs in kink interactions, where the kinks acquire internal structure. 

\section*{Acknowledgements}

AM acknowledges financial support from CNPq (Conselho Nacional de Desenvolvimento Científico e Tecnológico), Grant no. 306295/2023-7, and CAPES (Coordenação de Aperfeiçoamento de Pessoal de Nível Superior). AM also acknowledges support from the Beaufort Visiting Fellowship at St John’s College, University of Cambridge. Part of the simulations exhibited here were performed in the supercomputer SDumont of the Brazilian agency LNCC (Laboratório Nacional de Computação Científica). 

\bibliographystyle{unsrt}
\bibliography{references}

\end{document}